\documentclass[notoc]{JHEP}
\usepackage{epsfig}
\title{Quantum Corrections to the Kinetic Term in the Randall-Sundrum model} 
\author{Ian G. Moss and Wade Naylor \\ Department of Physics,
University of Newcastle Upon Tyne,\\ NE1 7RU U.K.}
\author{Wade Naylor} 
\abstract{
The effective action of the radion in the Randall-Sundrum model is
analysed. Fine tunings are needed to obtain the observed mass
hierarchy and an invisible radion. Since the kinetic terms are important for
determining the radion mass, the finite quantum corrections from
massless conformally coupled fermions are analysed and found to vanish at
one loop order.
}  
\preprint{hep-th/0106228}
\def\case#1/#2{{\textstyle {#1\over #2}}}
\begin{document}

\section{Introduction}

Interest in the possibilities of large extra
dimensions \cite{ADD} and the solution to the hierarchy problem for
non-factorisable geometries
 \cite{RS}, has sparked a renewed effort in
Kaluza-Klein theories. In particular,
much work has been done on one
loop quantum effects in the Randall-Sundrum
model
\cite{GPT,GR,T,FT,FMT,brevik,ponton}, with the hope of stabilizing the radion
field. In \cite{GPT} it was shown that massless bulk fermions lead to
stability, but the parameters of the model have to be fine tuned to obtain the
observed mass hierarchy. It was also argued that the radion mass was smaller
than a $TeV$ and ruled out by experiment. This was also confirmed for massive
bulk fermions \cite{FMT}. Consequently, classical  or non-perturbative
mechanisms have been saught to stablise the radion \cite{GW}.

In fact, we shall argue below that the radion mass can be consistent with
experiment if there is one additional fine tuned parameter in the model. This
makes a total of three parameters to be fine tuned, as opposed to just two in
this context (the cosmological constant and the TeV Higgs mass) in the
standard model. However, it has long been a mystery why the effects of quantum
gravity do not introduce Planck scale corrections into the standard model.

In the earlier work on Kaluza Klein theories \cite{DJT,ACF,CW} it was found
that the kinetic terms in the effective action played an important role.
Although they have no effect on the size of the internal dimension, they can
lead to instability in what would otherwise be a stable compactification. 

In this paper we calculate the finite one loop correction to the radion
kinetic term from massless fermions and find that it vanishes to second order
in derivatives. The calculation adapts a diagramatic method for obtaining
derivative expansions of a heat kernel \cite{MN,moss}, which is closely related
to the `covariant perturbation theory' first devised by Vilkovisky
\cite{osborn,vil,bar,avr}. In this case the fine tunings are protected at one
loop order.

\section{The radion action}

We will give a simplified picture of how one obtains the radion field from the
five dimensional action (along the lines of \cite{KOST}). The fifth dimension
will be taken to be an orbifold $S^1/Z_2$, then the full set of classical field
equations can be obtained from an action 
\begin{equation} 
S=-{1\over8\pi G_5} \int_{\cal M}\left(R-2\Lambda\right)
-{1\over4\pi G_5} \int_{\partial\cal M}K
- \int_{\partial\cal M}{\cal L}_m,
\end{equation}
where the boundary ${\partial\cal M}$ consists of a hidden brane $\Sigma_h$
and a visible brane $\Sigma_v$ with extrinsic curvature scalars $K$. The 
Lagrangian density ${\cal L}_m$ represents brane matter fields, and the
action correctly reproduces the brane boundary conditions \cite{chamblin}. We
shall be mostly concerned with the vacuum energies ${\cal V}_h$ and ${\cal
V}_v$. In the original Randall-Sundrum model, ${\cal V}_h=-{\cal V}_v={\cal
V}_0$, where 
\begin{equation} 
{\cal V}_0={3\kappa\over 4\pi G_5}
\end{equation}
and $\Lambda=-6\kappa^2$. 

We will identify the radion with the relative motion of the branes on a fixed
Randal-Sundrum background
\begin{equation} 
	ds^{2}=a(y)^2g_{\mu\nu}dx^\mu dx^\nu+dy^{2}, 
    \label{met1} 
\end{equation}
with $a(y)=\exp(-\kappa y)$ and Ricci scalar $R=-20\kappa^2$. This assumes
that the brane motion and the five dimensional gravity waves decouple at
leading order.

The moving branes are located at $y=y_h(x)$ and $y=y_v(x)$, with extrinsic
curvatures given by $K_h=-K_v$, 
\begin{equation}
K_v|g|^{1/2}\approx-4\kappa a_v^4-a_v\,\partial^2 a_v
-\kappa a_v^2(\partial a_v)^2
\end{equation}
where $\partial^2$ is the d'Alembertian operator. Substitution into the
original action gives a reduced Lagrangian ${\cal L}=T-V$, where
\begin{eqnarray}
T&=& \case1/2C_v \kappa^2(\partial a_v)^2-
\case1/2C_h \kappa^2(\partial a_h)^2\\
V&=&B_v\kappa^4a_v^4-B_h\kappa^4 a_h^4
\end{eqnarray}
and
\begin{eqnarray}
B_v\kappa^4&=&{\cal V}_0+{\cal V}_v\\
B_h\kappa^4&=&{\cal V}_0-{\cal V}_h\\
C_v\kappa^4&=&2{\cal V}_0+{\cal V}_v\\
C_h\kappa^4&=&2{\cal V}_0-{\cal V}_h
\end{eqnarray}
Provided that the motions are small, the reduced action will generate a
consistent set of field equations. The equations are incomplete because the 
metric variations have been restricted, but for spatially homogenous fields
the field equations can be completed with a single constraint $T+V=0$
\cite{KOST}.

The relative motion of the two branes can be isolated by introducing a change
of variables,
\begin{eqnarray}
\sigma&=&y_v-y_h\\
a&=&C^{-1}\left(C_ha_h^2-C_va_v^2\right)^{1/2}
\end{eqnarray}
where $C=(C_hC_v)^{1/2}$. In the new variables,
\begin{eqnarray}
T&=&\case1/2Z_T\kappa^4(\partial\sigma)^2
-\case1/2C\kappa^{-2}(\partial a)^2\\
V&=&Z_V\left(B_h-B_ve^{-4\kappa\sigma}\right)
\end{eqnarray}
where
\begin{eqnarray}
Z_T&=&2C^3a^2(C_he^{\kappa\sigma}-C_ve^{-\kappa\sigma})^{-2}\\
Z_V&=&C^2a^4e^{2\kappa\sigma}(C_he^{\kappa\sigma}-C_ve^{-\kappa\sigma})^{-1}
\end{eqnarray}
The negative kinetic term is associated with the fact that $a$
represents a gravitational degree of freedom of the double brane system, and
corresponds to a Friedmann equation of the usual form. The Friedmann
equation in $a_v$, which still determines the expansion rate of the visible
brane, has non-standard signs leading to doubts about the
consistency of the Randall-Sundrum model and standard cosmology \cite{mennim}.
The crucial idea here is that when the two branes are tied together by Casimir
forces, the dynamical equations are simplest when expressed in terms of the
collective coordinate $a$ and predict the usual cosmological evolution.

The classical theory has the shortcoming that the potential does not have a
minimum. This has lead to the consideration of one loop effects \cite{GPT}.
Calculations of the vacuum energy for $g_f$ massless fermion fields leads to a
potential
\begin{equation}
V=Z_V\left(B_h-B_ve^{-4\kappa\sigma}+
Ae^{-4\kappa\sigma}(1-e^{-\kappa\sigma})^{-4}\right)
\end{equation}
where $A$ is given in terms of the Riemann zeta function,
\begin{equation}
A={3\zeta(5)g_f\over 128\pi^2}.
\end{equation}
An equilibrium configuration requires both $V=0$ and $V'=0$. This implies
that $B_v>A$ and gives a relationship between $B_h$ and $B_v$

The ratio of the mass scale and the Planck scale on the brane is set by
$\lambda=e^{-\kappa \sigma}$. In order to obtain a mass hierarchy close to the
measured value we need the value of $B_v$ to lie very close to the value of
$A$. The mass of the radion (to a relative accuracy of $\lambda$) is then
\begin{equation}
m_\sigma^2=AC^{-1}\kappa^{-2}\lambda^3
\end{equation}
As Garriga et al. \cite{GPT} have pointed out, a radion with this mass should
have been observed in particle experiments before now. However, at the expense
of one further fine tuned parameter, the radion can be invisible with a
large mass for small values of $C_v$, which corresponds to
selected ranges of ${\cal V}_0$. A consistent radion model requires three fine
tuned parameters,  
\begin{eqnarray}
{\cal V}_v&\approx& -2\kappa^4A\\
{\cal V}_h&\approx& \kappa^4A\\
{\cal V}_0&\approx& \kappa^4A
\end{eqnarray}
We will not attempt to explain why the Lagrangian should have such fine tuned
parameters. However, the importance of the kinetic terms in fixing the radion
mass has lead us to examine the one loop corrections to these terms in order
to test the robustness of the fine tunings at the one loop level.
  
\section{The Basic Method}

We will obtain a perturbative expansion of the one loop effective action for
massless fermions on the moving brane background. The coordinate system is
chosen so that the coordinate in the fifth dimension is constant on the
boundaries, and then an expansion in derivatives of the metric is performed.
An analysis of the different fermion boundary conditions can be found in
\cite{FMT}. We shall consider the case of fermion components which satisfy
Neumann boundary conditions, but the other possibilities can be treated in an
equivalent way.

The conformal invariance of the massless fermions allows us to simplify the
problem by using the conformally related flat background metric,
\begin{equation}
ds^{2}=dx^{2}+d\tau^{2}.\label{met2}
\end{equation}
with $\tau_h(x)<\tau<\tau_v(x)$. We can replace $\tau$ with the
coordinate $\theta$ which is constant on the boundaries,
\begin{equation}
\theta=\pi(\tau-\tau_h(x))/\beta(x)
\end{equation}
where $\beta(x)=\tau_v(x)-\tau_h(x)$. The metric becomes  
\begin{equation} 
    ds^{2}=dx^{2}+\beta^{2}\pi^{-2}(d\theta+A_{\mu}dx^{\mu})^{2},
\label{pmetric}
\end{equation} 
where  
\begin{equation} 
A_{\mu}=\theta_c a_\mu+\case\pi/2 b_\mu 
\label{Amu} 
\end{equation}
for $\theta_c=\theta-\case\pi/2$ and
\begin{eqnarray}
a_\mu&=&\beta^{-1}(\partial_{\mu}\beta)\\
b_\mu&=&\beta^{-1}\partial_{\mu}(\tau_h+\tau_v)
\end{eqnarray}
The unperturbed configuration is therefore $A=0$ and $\beta$ constant.

The one loop effective action $W$ can be related to the Laplacian $\Delta$ of
the metric (\ref{pmetric}). We shall obtain a derivative expansion of $W$ using
heat kernel techniques. The heat kernel $K(x,\theta,x',\theta',t)$ satisfies
the equation  
\begin{equation} 
{\partial K\over \partial t}-\Delta K=
\delta(x-x')\delta(\theta-\theta')\delta(t)\label{hk} 
\end{equation}
The explicit form of the Laplacian is
\begin{equation} 
	\Delta=-\pi\beta^{-1}(\partial_{\mu}-\partial_{\theta}A_{\mu}) 
	\beta\pi^{-1}(\partial^{\mu}-A^{\mu}\partial_{\theta}) 
	-\pi^{2}\beta^{-2}\partial_{\theta}^{2}=\Delta_{0}+\Delta_{I}. 
\end{equation} 
where $\Delta_{I}$ denotes terms depending on derivatives of $\beta$,
$\Delta_1$ with one derivative, $\Delta_2$ with two derivatives etc. The first
terms are
\begin{eqnarray}
\Delta_{0} & = &  
	-\partial_{\mu}^{2}-\pi^{2}\beta^{-2}\partial_{\theta}^{2}, 
	 \\
\Delta_{1} & = &
	-\xi^{\mu}\pi^{2}(\partial_{\mu}\beta^{-2}) 
	\partial_{\theta}^{2}+A_{\mu}\partial_{\theta}\partial^{\mu} 
	+A^{\mu}\partial_{\mu}\partial_{\theta}, 
	 \label{pert}
\end{eqnarray}
where $\xi=x^{\prime}-x$.

It is possible to show using perturbation theory \cite{MN} that
the heat kernel has a representation in the form of a time ordered
exponential. In bra and ket notation,
\begin{equation}
K(x,\theta,x',\theta',t)=\,
\left\langle x'\theta'\left|e^{-\Delta_{0}t}\,T\exp\left[-\int_0^t 
e^{\Delta_{0}t'}\Delta_{I}e^{-\Delta_{0}t'}dt'\right]
\right|x\theta\right\rangle.  
\end{equation}
The calculation of the heat kernel can be performed most easily in momentum
space, therefore we introduce momentum basis states $|kn\rangle$ and vertex
operators 
\begin{equation}
V_{mn}\delta_{kk^{\prime}}=\langle k^{\prime}m|e^{\Delta_{0}t'}
\Delta_{I}e^{-\Delta_{0}t'}|kn\rangle.\label{vertex}
\end{equation}
For Neumann boundary conditions,
\begin{equation}
	  K(x,\theta,x,\theta,t)=\beta^{-1}\int \frac{d^{4}k}{(2\pi)^{4}}
	  \sum_{m,n} \cos\,m\theta\, \cos\,n\theta\,
e^{-\omega_{m}^{2}t-k^{2}t}\left(T\, \exp -\int
V(t^{\prime})dt^{\prime}\right)_{mn}
\end{equation}
The integrated trace of the heat kernel
\begin{equation}
K(t)=\int d^{4}x \int {d^{4}k\over(2\pi)^{4}}
	  \sum_{n=0}^\infty e^{-\omega_{n}^{2}t-k^{2}t}
	  \left(T\,\exp -\int V(t')dt'\right)_{nn}.
\end{equation}
As in reference \cite{MN}, we arrange the time ordered exponential in such a
way
that we can use Wick's theorem. Defining
\begin{equation}
	\langle ...\rangle=\frac{1}{Z(t)}\int \frac{d^{4}k}{(2\pi)^{4}}
	\sum_{n=0}^\infty e^{-\omega_{n}^{2}t}e^{-k^{2}t/2}...e^{-k^{2}t/2},
\end{equation}
with $Z(t)=(4\pi t)^{-2}\sum_{n} e^{-\omega_{n}^{2}t}$, gives
\begin{equation}
	  K(t)=\int d^{4}x Z(t) 
\left\langle T \exp\left(-e^{-k^{2}t/2}Ve^{k^{2}t/2}\right)
	  \right\rangle.
	  \label{htkrn}
\end{equation}
In order to construct the vertex operator we take the previous expresions for
$\Delta_I$ and make the replacements 
\begin{equation}
\xi^\mu\to\delta^\mu,\quad \partial_\mu\to\case1/2\delta'_\mu,\quad
\theta_c\to\overline\theta,\quad\partial_\theta^2\to -n^2
\end{equation}
where $\overline\theta(t')$ is defined in appendix B, $\delta_\mu'=-2ik_\mu$
and
\begin{equation}
\delta^{\mu}(t')=i\frac{\partial}{\partial k_{\mu}}
-2ik^{\mu}t^{\prime}+ik^{\mu}t.
\end{equation}
The expansion of the time ordered products leads to contractions between
the above operators which can be interpreted as propagators (see appendix A).

To begin with, consider $\tau_h+\tau_v=0$, i.e.
$A_{\mu}=\theta_ca_{\mu}$. 
If $A_{\mu}$ is substituted into (\ref{vertex}) one finds the vertex operators
\begin{eqnarray}
V_1&=& 2\omega_n^2\delta_{mn}a_\mu\delta^\mu+ 
\langle m|\overline{\theta}\partial_{\theta}|n\rangle 
a_\mu\delta^{\prime \mu}\\
V_2 &=& \omega_{n}^{2}\delta_{mn}(a_{\mu,\nu}
-2a_\mu a_\nu)\delta^{\mu}\delta^{\nu}
+\case1/2\langle m|\overline{\theta}\partial_\theta|n\rangle
\,a_{\nu,\mu}(\delta^\mu\delta^{\prime \nu}
+\delta^{\prime \nu} \delta^\mu)
-a^{2}\langle m|({\overline{\theta}}\partial_\theta)^2|n\rangle,
\end{eqnarray}
where $\omega_{n}^{2}=\pi^{2}n^{2}/\beta^2$.

The last term in the expression for $V_{2}$ can be removed by the vertex
elimination trick described in \cite{MN}. This allows terms to be
eliminated by modifying the contraction
\begin{equation}
\stackrel{\leftrightarrow}{D}(t_i,t_j)=2\delta(t_i-t_j)-2t^{-1}.\label{daa}
\end{equation}
The remaining vertex $V_2$, and all higher order vertices, can be obtained by
differentiating $V_1$.  

\EPSFIGURE[ht]{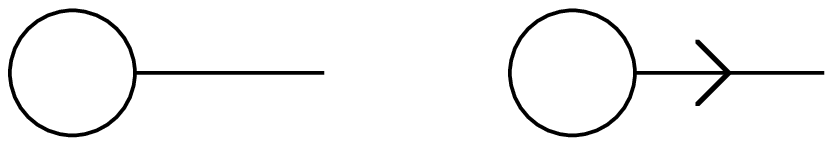}{Diagrams corresponding to the two terms in the vertex
operator $V_1$. The line represents the $\delta$ operator and the line with an
arrow represents the $\delta'$ operator.\label{fig1}}

\EPSFIGURE[ht]{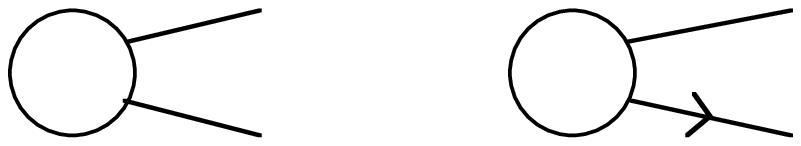}{Diagrams corresponding to the vertex operator
$V_2$. Each additional line represents an additional $\delta$ operator. 
\label{fig2}}

The vertex operators can be represented diagramatically by vertices with
external legs representing the $\delta$ operators (see figure \ref{fig1} and
figure \ref{fig2}). Each diagram contributes to the heat kernel (\ref{htkrn}),
or equivalently to the zeta function via a Mellin transform, 
\begin{equation}
\zeta_k(s)=\int d^4x
{1\over\Gamma(s)}\int t^{s-1} Z(t)\langle...\rangle
dt\label{zeta}.
\end{equation}
The one loop effective action is given by $W=\case1/2\zeta'(0)$.

\section{Second order calculation}

We now perform the calculation of the contribution to the kinetic terms to
second order.
 There are only three diagrams that contribute to the effective
action.
Other diagrams are zero due to the proper time integrals (see appendix
A) or theta integrals (see appendix B).

\EPSFIGURE[ht]{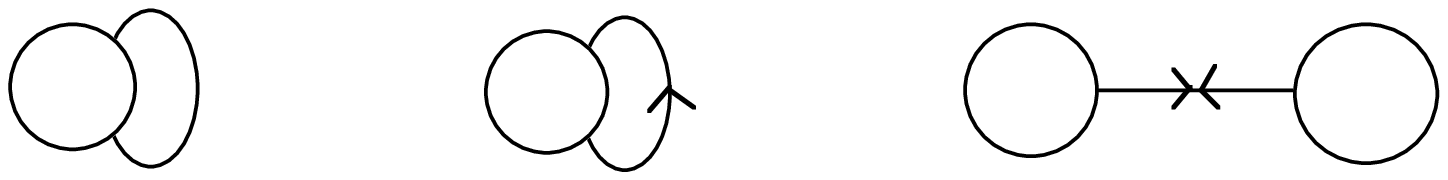}{The non-zero diagrams: (a) gives a total divergence; 
(b) and (c) cancel out.
\label{fig3}}

The first diagram is shown in figure \ref{fig3}a with one second
order vertex and one contraction $D(t_1,t_1)$.
This diagram gives a
contribution to the heat kernel of (see equation (\ref{htkrn})),  
\begin{equation}
Z(t)\langle (3a) \rangle =-\frac{1}{2}\frac{1}{(4\pi t)^2}
    (\partial^\mu a_\mu-2a^2)\frac{1}{3}t^2\sum_n e^{-\omega_n^2 t}\omega_n^2,
\end{equation}
where the first factor is a symmetry factor and the first integral in (A7) has
been used. The contribution of the diagram to
the zeta function
(denoted by $\zeta_a$) is given by equation (\ref{zeta}), 
\begin{equation}
\zeta_a(s)=-\frac{1}{16\pi^2}\int
d^4x\frac{1}{6}(\partial^\mu a_\mu-2a^2)\sum
\omega_n^{2-2s}. 
\end{equation}
The sum can be expressed in terms of the Riemann zeta function and
the derivative of the zeta function at $s=0$ is then
 
\begin{equation}
\zeta_a'(0)=-\frac{1}{16\pi^2}{\pi^2\over 3}\zeta_R'(-2)
\int d^4x\,\beta^{-2}(\partial^\mu a_\mu-2a^2).
\end{equation}
Using $a_\mu=\beta^{-1}\partial_\mu\beta$ and $\zeta_R'(-2)=
-\zeta_R(3)/(4\pi^2)$, we get
\begin{equation}
\zeta_a'(0)=-{1\over16\pi^2}{1\over48}\zeta_R(3)
\pi^2\int d^4x (\partial^2\beta^{-2}),\label{trm1}
\end{equation}
This term is a total divergence and makes no contribution 
unless the 3-branes have a boundary.

The second term corresponds to the diagram in figure \ref{fig3}b. 
Here
identities in appendix A and B have to be used. The diagram gives 
\begin{equation}
Z(t)\langle (3b)\rangle ={1\over(4\pi t)^2}
     \partial^\mu a_\mu\frac{1}{3}t^2\sum_n 
e^{-\omega_n^2t}\omega_n^2.
\end{equation}
The contribution to the effective action is
now
\begin{eqnarray}
\zeta_b'(0)&=&{1\over16\pi^2}{2\pi^2\over 3}\zeta_R'(-2)
\int d^4x\,\beta^{-2} (\partial^\mu a_\mu)\nonumber\\
&=&{1\over16\pi^2}{4\pi^2\over3}\zeta_R^{\prime}(-2)
\int d^4x \beta^{-4}(\partial\beta)^2,\label{trm2}
\end{eqnarray}
where we have used equation (\ref{Amu}) and integrated by
parts.

The third term corresponds to the diagram in figure \ref{fig3}c. For this
diagram
 
\begin{equation}
Z(t)\langle (3c) \rangle =-\frac{1}{2}\frac{1}{(4\pi t)^2}
     a^2 \frac{4}{3} t^2 \sum_n e^{-\omega_n^2 t}\omega_n^2.
\end{equation}
The contribution to the effective action is 
\begin{eqnarray}
	\zeta_c ^{\prime}(0) & = &-\frac{1}{16\pi^2}{4\pi^2\over 3}
\zeta_R^{\prime}(-2)\int d^4x\,\beta^{-2}a^2
	 \nonumber\\
	&=&-\frac{1}{16\pi^2}
{4\pi^2\over 3}\zeta_R^{\prime}(-2)
\int d^4x \beta^{-4}(\partial\beta)^2.
	\label{trm3} 
\end{eqnarray}
Thus we see that equations (\ref{trm2}) and (\ref{trm3})
cancel.
A similar
calculation shows that the terms involving $b_\mu$ also cancel.
\section{Conclusion}

The method which we have used to evaluate the kinetic terms in the radion
action for a brane world model can be extended quite easily. We have explicitly
shown that there are no one loop second order corrections from massless
fermions. It also follows by a simple extension of the arguments that there are
no one loop second order corrections from any conformally invariant bulk fields
on a conformally flat background or from massive fields on a flat background.
The results are therefore not restricted to the Randall-Sundrum model and
could be used for other brane world scenarios.

We might ask whether there are any derivative terms in the one loop
corrections to the radion action. In fact, they do have to exist because of the
renormalisation scale dependence. The renormalisation scale enters along with
the $a_5$ heat kernel coefficient \cite{branson}, which is of order $K^4$ in
the extrinsic curvature and therefore eighth order in derivatives. We do not
know yet whether this is the leading term in the derivative expansion. 

Although the calculation of one loop effects may have a bearing on the
hierarchy problem, the quantum corrections can also be considered in contexts
which are independent of the hierarchy problem. For example, the cosmological
evolution of these models appears to offer advantages over other
stabalisation mechanisms in the way that the expansion couples to the energy
density on the brane \cite{mennim}. It would be interesting to persue this
point in more detail.

\appendix

\section{Propagators}
 
There are three different propagators
corresponding to the distinct 
 combinations of the operators,
\begin{eqnarray}
 \langle T\delta^\mu(t_1)\delta^\nu(t_2)\rangle&=&
\delta^{\mu\nu}D(t_1,t_2)\\
\langle T\delta'_\mu(t_1)\delta^\nu(t_2)\rangle&=&
\delta_\mu{}^\nu\overrightarrow D(t_1,t_2)\\
\langle T\delta'_\mu(t_1)\delta'_\nu(t_2)\rangle&=&
\delta_{\mu\nu}\stackrel{\leftrightarrow}{D}(t_1,t_2).\label{aad}
\end{eqnarray}
These can be evaluated using the creation and annihilation operators
defined in \cite{MN}, leading to
\begin{eqnarray}
D(t_1,t_2)&=&
2\hbox{min}(t_1,t_2)-2t^{-1}t_1t_2\\
\overrightarrow D(t_1,t_2)&=&
2\theta(t_2-t_1)-2t^{-1}t_2
\end{eqnarray}
where $\theta(t)=1$ for $t\ge 0$ and zero otherwise.
The third
contraction evaluates to $-2t^{-1}$, but we replace (\ref{aad}) with
\begin{equation}
\stackrel{\leftrightarrow}{D}(t_1,t_2)=2\delta(t_1-t_2)-2t^{-1}
\end{equation}
in order to reduce the number of vertices. The relevant time integrals
of the propagators are: 
 
\begin{eqnarray}
	   \int_0^t D(t_1,t_1)dt_1=\frac{1}{3}t^2 &\hspace{1.5cm}& 
	   \int_0^t D(t_1,t_2)dt_1dt_2=\frac{1}{6}t^3
	 \nonumber \\
       \int_0^t \overrightarrow D(t_1,t_2)dt_1dt_2=0 &\hspace{1.5cm}& 
       \int_0^t t_1\overrightarrow D(t_1,t_2)dt_1dt_2=0  
	 \nonumber\\
	\int_0^t(\overrightarrow D(t_1,t_1)+\overleftarrow D(t_1,t_1)) dt_1=0 &&
	\int_0^t t_1(\overrightarrow D(t_1,t_1)+\overleftarrow D(t_1,t_1)) 
	              dt_1=-\frac{1}{3}t^2
	 \nonumber\\
	   \int_0^t \stackrel{\leftrightarrow}{D}(t_1,t_2)dt_1dt_2=0 &\hspace{1.5cm}&
	   \int_0^t (t_1+t_2) \stackrel{\leftrightarrow}{D}(t_1,t_2)dt_1dt_2=0
           \nonumber\\
	   \int_0^t Max(t_1,t_2)\stackrel{\leftrightarrow}{D}(t_1,t_2)dt_1dt_2
	   =-\frac{1}{3}t^2&\hspace{1.5cm}&
	   \int_0^t t_1t_2\stackrel{\leftrightarrow}{D}(t_1,t_2)dt_1dt_2 
	   =\frac{1}{6}t^3 
\end{eqnarray}

\section{Theta integrals}

The $\theta$ direction is special because the modes in this coordinate are
discrete, $\langle\theta|n\rangle=\sqrt{2}\cos(n\theta)$ for Neumann boundary
conditions and $\langle\theta|n\rangle=\sqrt{2}\sin(n\theta)$ for Dirichlet
boundary conditions. We need a way to replace a $\theta$ sandwiched between
$e^{\Delta_0t}$ terms inside the $|kn\rangle$ matrix elements. For this we
use the Cambell-Baker-Hausdorff formula,
\begin{equation}
e^{\Delta_0t}\,\theta_c\,e^{-\Delta_0t}=\overline{\theta}(t).  
\end{equation}
where $\overline{\theta}(t)$ is an operator
\begin{equation}
\overline{\theta}(t)=\theta_c-2\pi^2\beta^{-2}t\partial_\theta
\end{equation}
An important contraction identity we require for figure 3c is
\begin{equation}
\langle n|T\overline\theta(t_1)\partial_\theta
\overline\theta(t_2)\partial_\theta|n\rangle=
-\case1/{12}\beta^2\omega_n^2+4\omega_n^2{\rm max}(t_1,t_2)
\pm\omega_n^2(t_1+t_2)+4\omega_n^4t_1t_2.
\end{equation}
The results for other $\theta$ dependent terms in the matrix elements are given
by
 
\begin{eqnarray}
\langle n|n\rangle=1 &\hspace{2cm}&\langle n|\partial_\theta|n\rangle=0
\nonumber\\
\langle n|\theta_c\partial_\theta|n\rangle=\pm{1\over2}
&\hspace{2cm}&\langle n|\theta_c^2|n\rangle={\pi^2\over12}
\nonumber\\
\langle n|(\theta_c\partial_\theta)^2|n\rangle=-{\pi^2 n^2\over12}
&\hspace{2cm}&\langle n|\partial_\theta^2|n\rangle=-n^2.
\end{eqnarray}
The upper signs correspond to Neumann boundary conditions and the lower to
Dirichlet boundary conditions.



\end{document}